\documentclass[format=acmsmall, review=false, screen=true]{acmart}

\usepackage{tabularx,ragged2e,booktabs}
\usepackage{todonotes}

\acmJournal{TMIS}
\acmVolume{10}
\acmNumber{4}
\acmYear{2019}
\acmMonth{12}
\copyrightyear{2019}

\setcopyright{none}
\acmPrice{}

\acmDOI{10.1145/3370082}

\begin{document}

\title[Measuring the Business Value of Recommender Systems]{Measuring the Business Value of Recommender Systems}

\author{Dietmar Jannach}
\affiliation{
\institution{University of Klagenfurt}
\city{Klagenfurt}
\country{Austria}
}

\author{Michael Jugovac}
\affiliation{
\institution{TU Dortmund}
\city{ Dortmund}
\country{Germany}
}
\begin{abstract}
Recommender Systems are nowadays successfully used by all major web sites---from e-commerce to social media---to filter content and make suggestions in a personalized way. Academic research largely focuses on the value of recommenders for consumers, e.g., in terms of reduced information overload. To what extent and in which ways recommender systems create \emph{business value} is, however, much less clear, and the literature on the topic is scattered.
In this research commentary, we review existing publications on field tests of recommender systems and report which business-related performance measures were used in such real-world deployments. We summarize common challenges of measuring the business value in practice and critically discuss the value of algorithmic improvements and offline experiments as commonly done in academic environments. Overall, our review indicates that various open questions remain both regarding the realistic quantification of the business effects of recommenders and the performance assessment of recommendation algorithms in~academia.

\end{abstract}




\maketitle

\section{Introduction}
Recommender systems
are among the most visible and successful applications of Artificial Intelligence and Machine Learning technology in practice.
Nowadays, such systems accompany us through our daily online lives---for example on e-commerce sites, on media streaming platforms, or  in social networks. They help us by suggesting
things that are assumed to be of interest
to us and which we are correspondingly likely to inspect, consume, or purchase.

Recommendations that are provided online are usually designed to serve a certain purpose and to create a certain value, either for the consumer, the provider, some other stakeholder like an item producer, or several of them in parallel \cite{JannachAdomavicius2016,Abdollahpouri:2017:RSM:3079628.3079657,abdollahpouri2019personalization}. Academic research mostly focuses on the consumer perspective, with the implicit assumption that improved customer value is indirectly also beneficial for the recommendation provider. Indeed, among other considerations, service providers are usually interested in improving the recommendation experience of consumers. Typically, however, they assess the value of a recommendation system more directly in terms of \emph{business} value. Relevant business measures in that context include sales or revenue, click-through rates (CTR), higher user engagement, or customer retention rates \cite{Gomez-Uribe:2015:NRS:2869770.2843948,Garcin:2014:OOE:2645710.2645745,DBLP:conf/aaai/SmythCO07,JannachHegelich2009}.

Given the insights from the literature \cite{NetflixBlog2012,Davidson:2010:YVR:1864708.1864770,Gomez-Uribe:2015:NRS:2869770.2843948,KatukuriEbay2014,Katukuri2015RIR}, it is undisputed that recommender systems can have positive business effects in a variety of ways. However, \emph{how large} these effects actually are---compared to a situation without a recommender system or with a different algorithm---is not always clear. In the literature, the reported numbers vary largely, from marginal effects in terms of revenue \cite{Dias:2008:VPR:1454008.1454054} to orders of magnitude of improvement in terms of ``Gross Merchandise Volume'' \cite{Chen:2011:REI:2009916.2010051}. Furthermore, in some application domains, it might also not be immediately evident what particular measure one should focus on.
Increases in click-through rates are, for example, often used as a measure in reports on real-world deployments. To what extent such increases actually reflect the long-term business value of a recommender, can, however, be open to question.

A related challenge---in theory as in practice---is to predict if a planned improvement of the used recommendation algorithm will positively affect a certain business measure. In fact, many companies are constantly trying to improve their recommendation systems, and they usually run field tests (A/B tests) to gauge the effects of certain changes. Since such field tests can be costly and risky, companies like Netflix additionally rely on \emph{offline} experiments based on historical data to make preliminary assessments of planned algorithm changes \cite{Gomez-Uribe:2015:NRS:2869770.2843948}. This type of experiment is also predominant in the academic literature, mostly because researchers typically have no access to a real-world system where they can test the effectiveness of their ideas. Unfortunately, while nowadays a number of research datasets are available, they usually do not contain quantitative data from which the business value can be directly inferred. Furthermore, since the choice of a business measures is often specific for a domain, researchers typically abstract from these domain specifics and, most commonly, aim at predicting user preference statements (e.g. ratings) or the user's next action as recorded in a given dataset. To what extent such measurements---and offline experiments in general---are helpful to assess the potential \emph{business value} of algorithmic improvements, is also open to question. According to a report by Netflix researchers \cite{Gomez-Uribe:2015:NRS:2869770.2843948}, offline experiments were \emph{not} found ``to be as highly predictive of A/B test outcomes as we would like.''

Overall, given the largely varying results reported on the effect of recommenders on business, two potential pitfalls can be identified: First, the business value of the recommender systems is not adequately defined, measured, or analyzed, potentially leading to wrong conclusions about the true impact of the system. Second, the value of deploying complex algorithms that are slightly better than previous ones in terms of an abstract computational measure like the RMSE might be wrongly estimated. After the Netflix Prize competition, for example, the winning strategy was never put into practice. Despite the theoretical accuracy gains, it was not clear if the potentially resulting increases in business value would justify the engineering effort to implement the winning strategy in a scalable manner \cite{NetflixBlog2012}.

In this research commentary, we therefore review the literature on \emph{real-world deployments} of recommender systems.
We consider both personalized recommendation approaches based on long-term user profiles as well as recommendations that are based on interactively acquired preferences or the user's current navigational context\footnote{Related-item recommendations like Amazon's ``Customers who bought {\ldots} also bought'' are an example of adaptive suggestions that are mainly based on the navigational context.}.
This review shall serve online service providers and retailers as a basis to assess the potential value of investing (more) into recommender systems technology. We will furthermore summarize the outcomes of scientific studies which aim to assess to what extent algorithmic improvements in terms of prediction accuracy lead to a better quality perception or higher adoption by users. Finally, we discuss possible implications of our survey for industry and academia.

\section{What We Know About the Business Value of Recommenders}
\label{sec:business-value}

\subsection{General Success Stories}
Companies usually do not publicly share the exact details about how they profit from the use of recommendation technology and how frequently recommendations are adopted by their customers. Certain indications are, however, provided sometimes in the form of research papers or blog posts. Netflix, for example, disclosed in a blog post \cite{NetflixBlog2012} that ``\emph{75\,\% of what people watch is from some sort of recommendation}'', and YouTube reports that 60\,\% of the clicks on the home screen are on the recommendations \cite{Davidson:2010:YVR:1864708.1864770}.
In another, later report on the system designed at Netflix \cite{Gomez-Uribe:2015:NRS:2869770.2843948}, the authors reveal that recommendations led to a measurable increase in user engagement and that the personalization and recommendation service helped to decrease customer churn by several percentage points over the years. As a result, they estimate the business value of recommendation and personalization as more than 1 billion US dollars per year.\footnote{How these numbers were estimated is, unfortunately, not specified in more detail. The total revenue of Netflix in 2017 was at about 11.7 billion dollars, with a net profit of 186 million dollars.} Another number that is often reported in online media\footnote{See, e.g., https://www.mckinsey.com/industries/retail/our-insights/how-retailers-can-keep-up-with-consumers}, is that according to a statement of Amazon's CEO in 2006, about 35\,\% of their sales originate from cross-sales (i.e., recommendation).

These examples illustrate the huge potential of personalization and recommendation. The fact that many successful online sites devote a major part of the available screen space to recommendations (e.g., the top screen area on YouTube \cite{Davidson:2010:YVR:1864708.1864770} and almost the entire screen at Netflix) is another indicator of the substantial underlying business value.  In the following, we summarize findings that were reported in the literature to draw a more detailed picture, e.g., in terms of the business value that recommendations achieve and how this is measured. Since not all forms of measurements might be similarly useful for each domain, we will later on discuss %differences that we identified in terms of how the measurements were taken, highlighting
why it is important to interpret the observed findings with care.

%differences in the testing methods and metrics, highlighting that not all measures are equally useful in estimating business value.

\subsection{What is Being Measured and What are the Reported Effects?}
\begin{figure}[h!t]
  \centering
  \includegraphics[width=1\linewidth]{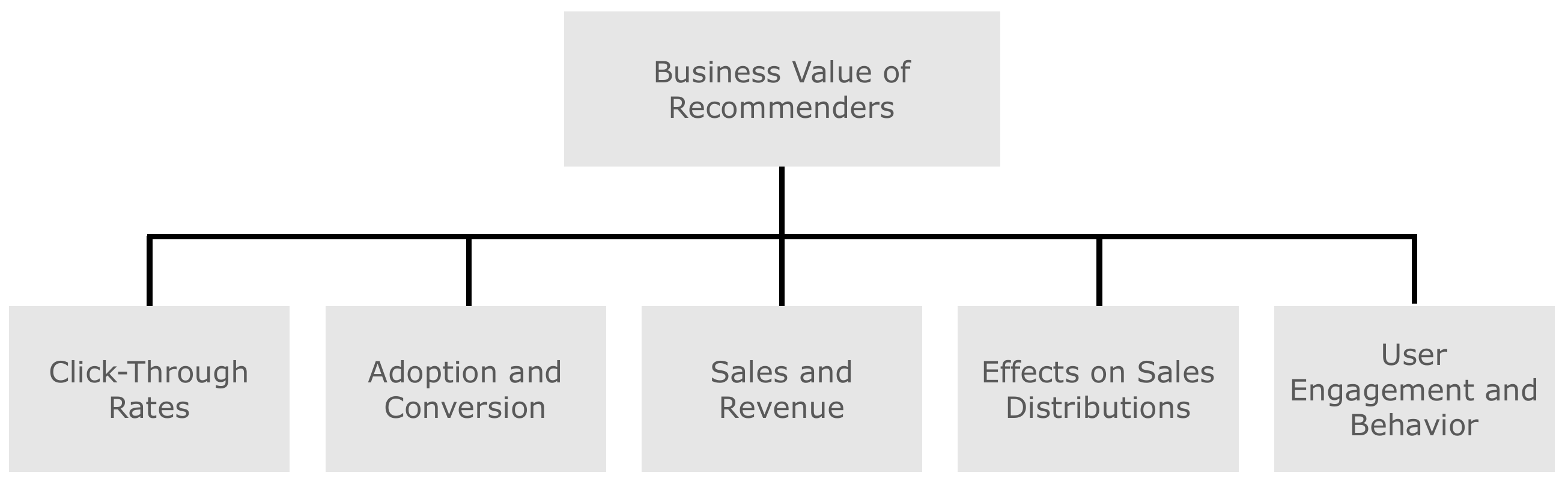}
  \caption{Overview of Measurement Approaches}
  \label{fig:taxonomy-of-measurement-approaches}
\end{figure}

The way companies measure the effects and the business value of a deployed recommender system depends on various factors, including the application domain and, more importantly, the business model of the company. Such business models can, for example, be partly or almost fully based on ads (e.g., YouTube or news aggregation sites). In this case, the goal might be to increase the time users spend with the service. Increased engagement is also often an objective when businesses (e.g., music streaming services) have a flat-rate subscription model, because engagement is often considered as a proxy for retention. In other domains, the effect of a recommender can be more direct (e.g., on an e-commerce site). The corresponding objectives are to either directly promote sales through the recommender or to obtain shifts in consumers' shopping behavior, e.g., towards more profitable items. In all of the aforementioned cases, the underlying business models and objectives determine how companies measure the value of a recommender.
Figure \ref{fig:taxonomy-of-measurement-approaches} shows an overview of the main measurement approaches found in the literature, which we discuss in more detail in the following sections.

\subsubsection{Click-through rates}
With the click-through rate (CTR), we measure in some form how many clicks are garnered by the recommendations.
The underlying assumption is that more clicks on the recommended items indicate that the recommendations were more relevant for the users.

The CTR is a very common measure in \emph{news recommendation}. In an early paper by Das et al.~\cite{Das:2007:GoogleNews} on Google's news personalization engine, the authors found that personalized recommendations led to an average increase in clicks of 38\,\% compared with a baseline that only recommends popular items. On some days, however, in case of highly attractive celebrity news, the baseline actually performed better. Different personalization algorithms were tested, but no clear winner was identified.

Kirshenbaum et al.~\cite{KirshenbaumForbes} later on reported on a live experiment at Forbes.com. The best-performing method in their live study was a hybrid content-based, collaborative system, leading to a 37\,\% increase in CTR over a time-decayed popularity-based baseline. Interestingly, this trivial popularity-based baseline was among the best methods in their live trial.

In ~\cite{Liu:2010:PNR:1719970.1719976}, Liu et al.~also experimented with a content-based, collaborative hybrid for Google News recommendations. One particularity of their method was that it considered ``local trends'' and thus the recent popularity of the items. According to their experiments based on live traffic on the site, considering local trends helped to increase the CTR compared to the existing method \cite{Das:2007:GoogleNews} by around 30\,\% for a subgroup of relatively active users.
However, the experiments also showed that the improved recommendations ``stole'' clicks from other parts of the Google News page and the algorithmic improvement did thus not lead to more clicks on the overall site.

Instead of considering only community trends, Garcin et al.~\cite{Garcin:2014:OOE:2645710.2645745} specifically consider the recent interests of individual, \emph{anonymous users} in the recommendation process.\footnote{This setting corresponds to a session-based recommendation scenario \cite{QuadranaetalCSUR2018}.} They compared their \emph{Context Tree} (CT) method both with a random recommender and one that suggests the most popular items. Interestingly, the random recommender was better in terms of the CTR than the ``Most Popular'' recommender.
The CT recommender turned out to be beneficial mostly for longer user sessions, where it led to a CTR increase of
about 35\,\%.

Besides news, the CTR was also used in a number of other application domains, including research article recommendation \cite{Beel2017TFIDu-41879,DBLP:conf/ercimdl/BeelL15,Beel:2013:CAO:2532508.2532511}, community recommendation on social networks \cite{Spertus:2005:ESM:1081870.1081956}, or video recommendation on YouTube \cite{Davidson:2010:YVR:1864708.1864770}. In the latter case, the authors report an increase of over 200\,\% in terms of the CTR when they used a comparably simple algorithm based on co-visitation instead of an approach that recommends the most viewed items.

In the special problem setting of ``similar item recommendations'', i.e., the recommendation of items in the context of a reference item, researchers at eBay have repeatedly reported on CTR improvements that were obtained through better algorithms \cite{Chen:2011:REI:2009916.2010051,KatukuriEbay2014,Katukuri2015RIR,Brovman:2016:OSI:2959100.2959166}. In \cite{KatukuriEbay2014}, for example, a 38\,\% CTR increase was observed in comparison to a simple title-based recommendation method; in \cite{Katukuri2015RIR}, a 36\,\% improvement in terms of the CTR was obtained for the ``related-item recommendations'' at the post-purchase page at eBay via a co-purchase mining approach. In \cite{Brovman:2016:OSI:2959100.2959166}, finally, only a minor increase in CTR (of about 3\,\%) was observed when applying a novel ranking method instead of a manually tuned linear model. Nevertheless, the model led to stronger increases in revenue (6\,\%) in the test period.

\subsubsection{Adoption and Conversion Rates}
Differently from online business models based on advertisements, click-through rates are typically not the ultimate success measure to target in recommendation scenarios. While the CTR is able to measure user attention or interest, it cannot convey, e.g., whether users really liked the recommended news article they clicked on or if they purchased an item whose product details they inspected based on a recommendation.

Therefore, alternative \emph{adoption} measures are often used that are supposed to be better suited to gauge the usefulness of the recommendations and which are often based on domain-specific considerations. YouTube, as reported in \cite{Davidson:2010:YVR:1864708.1864770}, uses the concept of ``long CTRs'', where clicks on recommendations are only counted if the user subsequently watched a certain fraction of a video~\cite{Davidson:2010:YVR:1864708.1864770}. Similarly, Netflix uses the ``take-rate'' as a measure which captures in how many cases a video or movie was actually played after being chosen from a recommendation \cite{Gomez-Uribe:2015:NRS:2869770.2843948}.
According to their experiments, increases of the \emph{take-rate} due to the deployment of a personalized strategy are substantial when compared to recommendations based on popularity. No detailed numbers are unfortunately reported in \cite{Davidson:2010:YVR:1864708.1864770} and \cite{Gomez-Uribe:2015:NRS:2869770.2843948} in that respect.

In domains where the items cannot be directly consumed (read, viewed, or listened to), other business-related adoption measures are common. Examples include the ``purchase-through'' or ``bid-through'' rate on eBay \cite{Chen:2011:REI:2009916.2010051}, as well as the ``link-through'' or ``cite-through'' rate for research paper recommendations \cite{DBLP:conf/ercimdl/BeelL15}, or the number of ``click-out'' events to external partners in online marketplaces \cite{Lerche2016}.

A longitudinal A/B testing phase of a new similar-item algorithm at eBay \cite{Chen:2011:REI:2009916.2010051}, for example, showed that the new system led to a bid-through rate between about 3.3\,\% and 9\,\%, depending on the product category. The purchase-through rate was at about 1.5\,\% and 3\,\%, measured at the same post-purchase (checkout) page. Overall, the authors conclude that their new system based on probabilistic clustering---if it would go live after six months of A/B testing and tuning---would lead to a 3-5 fold improvement over their current algorithm, which is a nearest-neighbor collaborative filtering method on the category level. In another field test at eBay \cite{KatukuriEbay2014}, the experimenters report an 89\,\% increase of ``add-to-wishlist'' actions after introducing a new similar-item recommendation algorithm on the page that users see after losing an auction compared to the previously used ``naive'' algorithm. On a different, much smaller marketplace for electronic gadgets, Lerche et al.~\cite{Lerche2016} found that using alternative recommendation strategies can increase the ``clickout'' rate to an external marketplace by more than 250\,\%.

In a quite different application domain, people-to-people recommendation on online dating portals, Wobke et al.~\cite{Adeployed2015} observed a significant increase in different domain-specific measures (e.g., ``10.9\,\% lift in positive contacts per user'' or ``4.8\,\% lift in open communications per user'') when a collaborative filtering method was applied instead of a baseline that matches explicit user profiles. In another people-to-people recommendation scenario, matchmaking on a job portal, the authors found that collaborative filtering helped improve their specific performance measure---the probability of a user contacting an employer after seeing a recommendation---by more than 200\,\% over a popularity-based baseline \cite{Adeployed2015}.
In the context of the LinkedIn platform, Bastian et al.~\cite{Bastian:2014:LSL:2645710.2645729} proposed a new method for \emph{skill} recommendations. A field test showed that recommending a list of skills to add to the profile led to a higher rate of users who added skills (49\,\% vs.~4\,\%) compared to a manual input system with type-ahead. Note, however, that two different user interface approaches were compared in this field test. It is thus not fully clear how much of the gains can be attributed to the recommendation method and how much can be explained by the more convenient way of adding skills.

The number of contact requests was also the success measure of the deployed tourism recommender system described by Zanker et al.~\cite{Zanker2008EvaluatingTourism}. In their quasi-experimental design, users who interacted with a conversational recommender were twice as likely to issue an accommodation request through the tourism website than users who did not. In this context, it is, however, important to keep in mind that the users who decided to use the recommender might have had a more specific interest than others when they entered the site. Also for the travel and tourism domain, Kiseleva et al.~\cite{Wheretogo2015} compared different strategies in a field test at Booking.com. Their experiment showed that a Naive Bayes ranking method led to a 4.4\,\% increase in conversion compared to the current online system. Note that in both mentioned studies from the tourism domain \cite{Zanker2008EvaluatingTourism,Wheretogo2015}, the recommendations are not based on long-term user profiles but on user preferences that are interactively collected before recommending.

\subsubsection{Sales and Revenue}
The adoption and conversion measures discussed in the previous section are, in many applications, more informative of the potential business value of a recommender than CTR measures alone. When users pick an item more often from a recommendation list which they later purchase or view, this is a good indicator that a new algorithm was successful to identify items that are relevant to the user. Nonetheless, it remains difficult to assess how such increases in adoption translate to increased business value. A recommender might, in fact, make many suggestions to users that they would purchase anyway (see \cite{BodapatiPurchaseData2008} for an analysis of this matter). The increase in business value might therefore be lower than what we can expect when looking at increases of adoption rates alone. Moreover, if the relevance of recommendations was very low already initially, i.e., almost nobody clicked on them, increasing the adoption rate even by 100\,\% might lead to very limited absolute extra value for the business.

Generally, there are various ``business models'' for recommenders, i.e., how they help improve a business. Chen et al.~\cite{Chen2009HICCS}, for example, list a number of effectiveness measures, including increased sales, fee-based sales through more transactions or subscriptions, and increased income from other types of fees, see also \cite{JannachAdomaviciusVAMS2017}. Unfortunately, only few papers report the effects of recommenders on such measures, partially because the data is confidential and partially because the effects cannot easily be isolated from each other.
In the case of Netflix, for example, renewed subscriptions are a desired effect of recommenders, but with very low churn rates in general it is difficult to attribute differences in churn rates to changes in a recommender algorithm \cite{Gomez-Uribe:2015:NRS:2869770.2843948}.

While today's video streaming sites, like Netflix, have flatrate subscription models, there are other sites where additional content can be purchased. Previously, such pay-per-view models were more common, and Bambini et al.~\cite{Bambini2011} investigate the business effects of a recommender for a \emph{video-on-demand} service. They not only measured what they call the ``empirical recall'', i.e., the fraction of recommended movies that were later on watched, but also tried to assess the additional video-on-demand sales induced by the recommender. However, because the recommender system was deployed to the whole user base instead of only a small treatment group, the authors had to gauge its performance by comparing the global number of video views in the weeks before and after the introduction of the recommender. They finally estimate the lift in sales obtained by their content-based approach to be 15.5\,\%, after smoothing out other factors such as marketing campaigns with a moving average.

Also in the media domain, Lee and Hosanagar investigate the impact of recommenders on the sales of DVDs of an online retailer \cite{DBLP:conf/icis/LeeH14}. They tested both purchase-based and view-based collaborative filtering approaches and observed a 35\,\% lift in sales when the purchase-based version was compared with a ``no recommendations'' condition. The increases in sales were much less pronounced (and not statistically significant) when a view-based strategy was employed.
Finally, they also observed that only recommending recently viewed items actually led to a slight decrease in overall sales. Differently from the findings in \cite{Lerche2016}, reminders were therefore not directly helpful in terms of business value. In real-world e-commerce applications, where such reminders are common \cite{JannachLudewigLerche2017umuai}, they might more often represent convenient navigation shortcuts for users than additional sales stimulants.

Besides the movie and DVD domains, a number of success stories of recommender systems exist for more general \emph{e-commerce settings}. One of the earliest reports that quantifies the effects of recommenders on business value focused on online grocery orders. In \cite{Lawrence2001}, the authors found through a pilot study that their revenue increased by 1.8\,\% through purchases that were made directly from the recommendation lists.

Dias et al.~\cite{Dias:2008:VPR:1454008.1454054} also evaluated a recommender for an online grocery store. They observed an increase in \emph{direct} revenue of only 0.3\,\% after deploying the system. However, they also discovered substantial \emph{indirect} effects, with increases of up to 26\,\% for one category. It, thus, became obvious that the recommender was able to inspire or stimulate additional sales even though consumers did not pick the items from a recommendation list. A similar effect was also reported in \cite{Lawrence2001}, where the authors observed that the grocery recommender successfully guided customers to product categories that they had not considered before. More recently, the authors of \cite{KamehkhoshJannachBonninMILC2018} also detected such an ``inspirational'' effect of recommender systems in the music domain.

In the context of similar item recommendations at eBay, the authors of \cite{Brovman:2016:OSI:2959100.2959166} report a 6\,\% improvement in terms of revenue when they field tested a novel method against a baseline linear model. Specifically, they proposed a two-stage approach, consisting of a candidate item retrieval and a subsequent ranking phase. The ranking model is based on logistic regression, where the purchase probabilities are computed based on observations from recommendations that were made in the past. While the reported improvement is significant, it is much lower than the one reported earlier in a similar context at eBay \cite{Chen:2011:REI:2009916.2010051}, where the authors observed an increase of the ``Gross Merchandise Bought'' measure of almost 500\,\% in the context of a specific part of the website. However, in general, it seems that such increases are only possible under certain circumstances, e.g., when the existing methods are not effective. The study reported in \cite{Chen:2011:REI:2009916.2010051} also only lasted one week and it is unclear if the system went into production.

An \emph{online book store} was the evaluation environment of another early work presented in \cite{Shani:2005:MRS:1046920.1088715}, where the authors compared two algorithmic approaches for next-item recommendations. The results showed that their new method led to 28\,\% more profit than when using a simple one; when they entirely removed the recommenders for one month, the revenue dropped by 17\,\%. However, the size of this effect could have also been influenced by additional factors like seasonal effects.

To what extent recommenders impact sales within an online marketplace for \emph{mobile phone games} was analyzed in \cite{JannachHegelich2009}. Here, the authors report the outcomes of a field test, where several algorithms were A/B tested for a number of weeks. The best method in their study, a content-based approach, led to an increase of sales of 3.6\,\% compared to the condition where no recommendations were provided. In the study, it turned out that the choice of the strategies should be made dependent on the user's navigational situation. While users might, for example, like content-based recommendation in general, these ``more-of-the-same'' suggestions are not helpful right after users have already purchased something. Therefore, even slightly higher increases in sales can be expected when the user's navigation context is considered.

Other field studies in the context of recommenders on mobile phones were also discussed in  \cite{Tam:2005:WPP:1245598.1245602} and \cite{DBLP:conf/aaai/SmythCO07}. Tam and Ho \cite{Tam:2005:WPP:1245598.1245602} found that personalized offers led to about 50\,\% more ringtone downloads compared to randomized offers. Smyth et al.~\cite{DBLP:conf/aaai/SmythCO07} observed a 50\,\% increase in user requests when the mobile portal was personalized, which in their case directly translated into revenue. The direct revenue boost through personalization was quantified for one provider as \$15 million per year.

\subsubsection{Effects on Sales Distributions}
The discussions so far clearly show that personalized recommendations can strongly influence the behavior of users, e.g., how many items they buy. This influence can, however, not only mean that \emph{more} items are bought, it might also result in the effect that \emph{different} items are bought, due to the persuasive potential of recommenders \cite{Yoo:2012:PRS:2380966}. Sellers might want to persuade customers to buy specific items for a variety of reasons. For example, to stimulate cross sales, recommendations can make customers aware of items from other categories that they might also be interested in or items that complement their previously purchased items. A clothes retailer might, for example, want to branch out into the shoes business, at which point customers can be recommended the matching pair of shoes for every pair of pants they buy. However, recommendations can also be used to persuade users to choose a premium item that offers a higher revenue margin for the seller instead of a low budget item to maximize per-category profits.

In \cite{persuasive2006}, for example, the introduction of an interactive recommender for premium cigars led to a significant shift in consumers' purchasing behavior. Specifically, the personalized recommendations led to more purchases in the long tail, and the sales spectrum was no longer dominated by a few topsellers. A shift of sales distributions introduced by recommenders was also noticed in an early work by Lawrence et al.~\cite{Lawrence2001} in an online supermarket application.

The distribution of what users consume is also a relevant measure at Netflix \cite{Gomez-Uribe:2015:NRS:2869770.2843948}. The key metric here is called ``Effective Catalog Size'' and expresses the amount of catalog exploration by users. An analysis shows that in the presence of personalized recommendations, this exploration tendency strongly increases, and a shift away from the most popular items is observed.
However, such a shift in the consumption distribution does not necessarily mean that there is more business value (e.g., more downloads, purchases, or clicks). In \cite{Liu:2010:PNR:1719970.1719976}, for example, an improved news recommender stole clicks from other parts of the website, i.e., there was no increase in overall user activity.

A recent analysis of the effects of recommenders on sales diversity can be found in \cite{DBLP:conf/icis/LeeH14} and \cite{LeeHosanagar2018}. The underlying question is whether recommenders help to promote items from the long tail or if they---in particular when based on collaborative filtering---rather help to boost sales of already popular items. To that purpose, the authors of \cite{LeeHosanagar2018} conducted a randomized field experiment on the website of a North-American online retailer. The study revealed that the presence of a recommender actually led to a \emph{decrease} in aggregate sales diversity, measured in terms of the Gini coefficient. While at the individual user level often more items in the catalog were explored, it turned out that similar users in the end explored the same kinds of products. Looking at niche items, recommender systems helped to increase item views and sales; but the increase of sales for popular products was even stronger, leading to a loss of market share of niche items.\footnote{A simulation-based analysis of concentration effects can be found in \cite{JannachLercheEtAl2015}. The analysis indicates that the choice of algorithm determines the strength and direction of the effects.}

\subsubsection{User Behavior and Engagement}
In various application domains, e.g., media streaming~\cite{Gomez-Uribe:2015:NRS:2869770.2843948}, higher user engagement is considered to lead to increased levels of user retention, which, in turn, often directly translates into business value. Increased user activity in the presence of a recommender is reported in a number of real-world studies of recommender systems. Various measures are applied, depending on the application domain.

In the news domain, for example, two studies \cite{Garcin:2014:OOE:2645710.2645745,KirshenbaumForbes} observed longer sessions when a recommender was in place. In \cite{Garcin:2014:OOE:2645710.2645745}, the visit lengths were 2.5 times higher when recommendations were shown on the page.
In the context of mobile content personalization, Smyth et al.~\cite{DBLP:conf/aaai/SmythCO07} report a 100\,\% increase in terms of user activity and more user sessions. For eBay's similar item recommendations, as discussed above, Katukuri et al.~\cite{KatukuriEbay2014} found that users were more engaged in terms of ``add-to-wishlist'' events. More user actions with respect to citations and links to papers were observed for the research paper recommender discussed in \cite{DBLP:conf/ercimdl/BeelL15}.

In the domain of music recommendation, Domingues et al.~\cite{Domingues2013} compared different recommendation strategies and found out that a recommendation strategy that combines usage and content data (called \emph{Mix}) not only led to higher acceptance rates but also to a 50\,\% higher activity level than the individual strategies in terms of playlist additions. The authors furthermore measured loyalty in terms of the fraction of returning users. They again found differences between the recommendation strategies and indications that acceptance rates, activity levels, and user loyalty are related.

In some papers, users activity is considered to be the most important performance measure. Spertus et al.~\cite{Spertus:2005:ESM:1081870.1081956}, for example, measured how many users of the social network Orkut actually joined one of the recommended communities. In the case of LinkedIn, the authors of \cite{Xu:2014:MPS:2623330.2623368} report that user engagement was strongly increased when a new recommender for similar profiles was introduced. Their activity measures included both profile views and email messages exchanged between recruiters and candidates.

For the particular case of the community-answering platform \emph{Yahoo! Answers}, Szpektor et al.~\cite{Szpektor:2013:REP:2488388.2488497} found that recommendations that were solely based on maximizing content similarity performed worse than a control group. However, after increasing the diversity of the recommendation lists, they observed a 17\,\% improvement in terms of the number of given answers and an increase of the daily session length by 10\,\%. These insights therefore support findings from existing research in the fields of Recommender Systems and Information Retrieval which stipulate that it can be insufficient to consider only the assumed relevance of individual items but not the diversity of the provided recommendation list as a whole \cite{Ziegler:2005:IRL:1060745.1060754,McNee:2006:AEA:1125451.1125659}.

\begin{comment}
\begin{table}
\begin{tabular}{|l|l|}
  \hline
  Measurement & Remarks \\
  \hline
  Profit and Revenue & xyz \\
  \hline
\end{tabular}
\end{table}
\end{comment}

\section{Discussion}
\label{sec:discussion}
\subsection{Challenges of Measuring the Business Value of Recommender Systems}
\subsubsection{Direct Measurements}
Our review in Section \ref{sec:business-value} shows that there are various types of effects of recommender systems that can be measured. In some application domains and in particular in e-commerce, the business value can be measured almost directly by tracking effects on sales or revenue that result from more sales or shifts in the sales distributions caused by the recommender.
In such cases, it is important to ensure that the choice of the measure is aligned with the business strategy. In some domains, increasing the sales volume (revenue) might be relatively easy to achieve by recommending currently discounted items \cite{JannachLudewigLerche2017umuai} or by promoting low-cost, high-volume items through the recommendations. % TODO: would be good to have a reference here.
This might, however, not always be the best business strategy, e.g., for retailers that want to promote premium products with high profit margins.

But even in cases where the business value can be directly measured, A/B tests are usually only conducted for a limited period of time, e.g., for a number of weeks. Such time-limited tests are not able to discover longitudinal effects. While a field test might indicate that promoting already popular items is more beneficial than promoting long-tail items \cite{LeeHosanagar2018,DBLP:conf/icis/LeeH14}, the recommendation of (at least some) long-tail items might have direct or indirect sales effects in the long run. Such effects can, for example, occur when customers discover additional item categories on a shop through the recommendations over time \cite{Dias:2008:VPR:1454008.1454054} or when customers later on switch to a paid version of a product that was originally recommended to them as a free trial \cite{JannachHegelich2009}.

\subsubsection{Indirect Measurements}
While click-through rates and certain forms of adoption rates measure in a direct way whether or not users click on the recommended items, they are---unless used in a pay-per-click scenario---in most cases not a true measure of business value. A high CTR for a news recommendation site might, for example, simply be achieved through clickbait, i.e., headlines that make users curious. In the long run, and possibly only some time after the A/B test, users might, however, not trust the recommendations anymore in case the items they clicked on were ultimately not relevant for them. Zheng et al.~\cite{Zheng:2010:CMR:1864708.1864759} investigate the problem of using CTRs as a success measure for recommendations on a media streaming site. Their analysis indicates that there can be a trade-off between the optimization of the CTR and the optimization of the ranking of the items according to their expected relevance for the users. As mentioned above, while recommending mostly popular items can lead to a higher CTR in many applications, such improvements are often an overestimation of the true value of the recommender \cite{BodapatiPurchaseData2008}.
Generally, such types of recommendations can also lead to \emph{feedback loops}, where the recommender system reinforces the popularity of already popular items. This might, in turn, lead to filter bubbles, a decrease in diversity, and a neglect of the potential of promoting items from the long tail \cite{LeeHosanagar2018,BorgesiusFilterBubble2016,Jiang2019DegenerateFL,JannachLercheEtAl2015}.

An overestimation of the value of a recommender system can also arise when using certain types of adoption rates. When nearly everything on a web page is personalized or some form of recommendation, e.g., in the case of Netflix, users are likely to choose whatever is recommended to them due to a mere presence effect~\cite{Koecher2019}. According to \cite{Amat:2018:APN:3240323.3241729}, Netflix is working on personalizing the covers (artwork) of their streaming content to persuade or convince users of its relevance for them. Counting only how often users start streaming such an item can therefore also be misleading as this measure would include users who started playing the movie but did not enjoy it in the end. Consequently, one has to decide carefully when to consider such a recommendation a success.

In the domain of mobile game recommendation, the authors of \cite{JannachHegelich2009} used click rates, conversion rates, and game downloads as business-related measures besides the sales volume. When comparing these measures, it turned out that neither item view nor even download counts were reliable predictors for the business success. Some recommendation algorithms, for example, raised consumer interest but did not lead to downloads. In terms of the download counts, it, furthermore, became evident that some algorithms had a bias to promote often-downloaded games that could be used in a free trial (demo). How often users later on also purchased the paid version and how this affected sales in the longer term was not clear from the experiment.

In several other domains, it can be even more difficult to assess the business value of a recommender. In the case of flat-rate subscription models for media streaming services, for example, \emph{user engagement} is typically considered to be correlated with customer retention. According to our discussions in Section \ref{sec:business-value}, there is strong evidence that recommenders have positive effects on the amount of user activity, e.g., in terms of session lengths or site visits.
In some cases, when customer retention is already high---like in the case of Netflix---obtaining significant improvements in customer retention can be difficult to achieve \cite{Gomez-Uribe:2015:NRS:2869770.2843948}. Depending on the domain, however, customer engagement can  be a viable proxy for the business value of a recommender.

Overall, we can identify a number of challenges when it comes to measuring the business value of recommender systems. Table \ref{tab:measurements-overview} shows a summary of some of our observations.

\begin{table}[h!t]
	\centering
	\caption{Measurements to Assess the Value of Recommenders.}
	\label{tab:measurements-overview}
	\small
	\begin{tabularx}{1\linewidth}{@{}lX@{}}
    \toprule
        Measurement & Remarks \\
    \midrule
    Click-Through Rates & Easy to measure and established, but often not the ultimate goal. \\
    Adoption and Conversion & Easy to measure, but often requires a domain- and application specific definition. Requires interpretation and does not always translate directly into business value.\\
    Sales and Revenue  & Most informative measure, but cannot always be determined directly.\\
    Effects on Sales Distribution & A very direct measurement; requires a thorough understanding of the effects of the shifts in sales distributions.\\
    User Engagement and Behavior & Often, a correspondence between user engagement and customer retention is assumed; still, it remains an approximation.\\

    \bottomrule
	\end{tabularx}
\end{table}

\subsection{Algorithm Choice and the Value of Algorithmic Improvements}
The reported improvements after deploying a new or modified recommender system vary largely according to our review in Section \ref{sec:business-value}. One of the reasons for this phenomenon lies in the \emph{baseline} with which the new system was compared. Sometimes, the improvements were obtained compared to a situation with no recommender \cite{Dias:2008:VPR:1454008.1454054,Lawrence2001}, sometimes the new system replaces a comparably simple or non-personalized (e.g., popularity-based) method \cite{Das:2007:GoogleNews,Shani:2005:MRS:1046920.1088715}, and, in a few cases, more elaborate strategies are compared to each other \cite{Shani:2005:MRS:1046920.1088715,JannachHegelich2009}.

In many cases where business effects are \emph{directly} measured, increases in sales between one and five percent are reported on average. The increases sometimes vary across different categories, e.g., more than 26\,\% for one category of an online grocery store. In one study \cite{Shani:2005:MRS:1046920.1088715},
the authors also report a 17\,\% \emph{drop} in sales when the recommendation component was removed for a week.
Overall, these numbers seem impressive, given that a lasting increase in sales of only 1\,\% or even less can represent a substantial win in absolute numbers for a large business.

In papers that rely on click-through rates, different forms of adoption rates, or domain-specific measures, we can also often observe substantial increases, e.g., a 200\,\% CTR increase over a trivial baseline at YouTube \cite{Davidson:2010:YVR:1864708.1864770}, a 40\,\% higher email response rate at LinkedIn \cite{Rodriguez:2012:MOO:2365952.2365961}, or an increase of the number of answers by 17\,\% at Yahoo! Answers \cite{Szpektor:2013:REP:2488388.2488497}.
To what extent these strong increases of indirect measurements translate into business value is, however, not always fully clear and often difficult to estimate.

What can generally be observed is that---in many papers---algorithms of different \emph{types} or \emph{families} are compared in A/B tests, e.g., a collaborative filtering method against a content-based method, or a personalized method against a non-personalized one. This was, for example, done in \cite{JannachHegelich2009} and the outcomes of this study show that the choice of the recommendation strategy (\emph{collaborative} vs.~\emph{content-based} vs.~\emph{non-personalized}) does matter both in terms of sales and in general user behavior.
Such studies are, however, different from many offline experiments conducted in academic research, which typically benchmark algorithms of similar type, e.g., different matrix factorization variants or sometimes even only different loss functions for the same learning approach. Whether the often tiny accuracy improvements reported in such offline experiments translate into relevant business value improvements when deployed in real-world environments remains difficult to assess, as published industrial field tests rarely focus on such fine-grained comparisons between similar algorithmic approaches.

Finally, a general limitation of the discussed works is that typical A/B tests reported here focus almost exclusively on the gains that can be obtained when different algorithms are used. The success of a recommender system can, however, be dependent on a number of other factors, including the users' trust in the recommender or the website as a whole \cite{Xiao:2007:EPR:2017327.2017335,NilashiJannachEtAl2016}, the perceived transparency of the recommendations, and, most importantly, the user interface. Garcin et al.~\cite{Garcin:2014:OOE:2645710.2645745}, for example, report a 35\,\% increase in CTR when they deployed a more sophisticated news recommendation method in a field test. However, at the end of the paper, they mention that they changed the position and size of the recommendation widget in an additional A/B test. This change, which was only at the presentation level, immediately led to an increase in CTR by 100\,\%. This suggests that at least in some applications, it seems more promising to focus both on the user experience and algorithmic improvements instead of investing only in better algorithms.

\subsection{The Pitfalls of Field Tests}
A/B tests, i.e., randomized controlled field tests, are usually considered the ultimate method of determining the effects on a user population caused by adding a recommender system to a website or improving an existing system, and large companies constantly test modifications to their service through such field tests \cite{Kohavi:2012:TOC:2339530.2339653}. A number of typical challenges of running such tests are discussed in \cite{Gomez-Uribe:2015:NRS:2869770.2843948} for the case of Netflix. In their case, A/B tests usually last for several months. The main metrics in their analyses are centered around customer retention and user engagement, which is assumed to be correlated with customer retention. To make sure that the observed differences of the retention metric are not just random effects, they apply statistical methods, e.g., to determine appropriate sample sizes.\footnote{See \cite{Siroker2013} about the statistics behind A/B tests.}
Despite this statistical approach, the authors of  \cite{Gomez-Uribe:2015:NRS:2869770.2843948} report that interpreting the outcomes of A/B tests is not always trivial. In case of unexpected effects, they sometimes repeated the tests to find that the effect did not occur again.\footnote{See also \cite{Kohavi:2012:TOC:2339530.2339653} for an analysis of surprising results of A/B tests at Microsoft Bing.}

Generally, running reliable A/B tests remains difficult, even for large companies like Google, Microsoft, or Amazon, for various reasons \cite{Deng:2013:ISO:2433396.2433413,Kohavi:2012:TOC:2339530.2339653}. One fundamental challenge lies in the choice of the evaluation criterion, where there can be diametrical objectives when short-term or long-term goals are considered. An extreme case of such a phenomenon is reported in \cite{Kohavi:2012:TOC:2339530.2339653} in the context of field testing Microsoft's Bing search engine. One of the long-term evaluation criteria determined at the executive level was the ``query share'', i.e., the relative number of queries served by Bing compared to the estimated overall market. Due to a bug in the system, the search quality went down significantly during a test. This, however, led to a strong short-term \textit{increase} in the number of distinct queries per user, as users needed more queries to find what they searched for. Clearly, in the long term, poor search quality will cause customers to switch to another search service provider. Similar effects can appear in the context of CTR optimization as discussed above.

Another more implementation-related challenge is that often large sample sizes are needed. Even small changes in revenue or customer retention, e.g., 0.5\,\%, can have a significant impact on business. Finding such effects with a certain confidence can, however, require a sample of millions of users. In many cases, it is also important to run tests for longer periods of time, e.g., several months, which slows innovation.
Furthermore, running A/B tests with existing users can also be risky, and some companies therefore limit tests mostly to new users \cite{Gomez-Uribe:2015:NRS:2869770.2843948}. Consider, for example, a music recommender system that in the past, mostly focused on relatively popular artists. A planned algorithmic change might aim at better \emph{discovery} support by recommending newer artists more frequently. Users who were acquainted with the existing system might notice this change, and as they are less familiar with the recommendations by the new system,  their quality perception might degrade, at least initially \cite{JannachLercheEtAl2015a}. However, this initially poor user response does not necessarily mean that the new system will not be accepted in the long run.

Existing research proposes a number of methods to deal with these challenges \cite{Deng:2013:ISO:2433396.2433413,Kohavi:2012:TOC:2339530.2339653,Chapelle:2012:LVA:2094072.2094078}, e.g., in the context of web search, but it is unclear if smaller or less-experienced companies implement such measures in their field tests. These problems of potentially unreliable or misleading test outcomes also apply for the research works that are reviewed above in Section \ref{sec:business-value}. In many of the discussed papers, the exact details of  the conducted A/B tests are not provided. Sometimes, authors report how long the test was run and how many users were involved. While in several cases the evaluation period lasts for several months, there are cases where certain tests were run only for a few days or weeks (after piloting and optimization) \cite{Chen:2011:REI:2009916.2010051,Szpektor:2013:REP:2488388.2488497,Liu:2010:PNR:1719970.1719976,JannachHegelich2009,LeeHosanagar2018}. Regarding sample sizes, often only a few hundred or around one thousand users were involved \cite{DBLP:conf/aaai/SmythCO07,Zanker2008EvaluatingTourism,Beel:2013:CAO:2532508.2532511}. In almost all surveyed cases, an analysis of the required sample size and detailed statistical analyses of the A/B tests were missing. It can therefore not always be concluded with certainty that the reported outcomes are based on large enough samples or that they are not influenced by short-term effects (see \cite{Kohavi:2012:TOC:2339530.2339653}).

\subsection{The Challenge of Predicting Business Success from Offline Experiments}
Given the complexity and cost of running field tests, the common approach in academic research is to conduct offline experiments on historical data. The most common evaluation method is to hide some of the available information from such a dataset and use the remaining data to learn a model to predict the hidden data, e.g., a user rating or other user actions like clicks, purchases, or streaming events.

This research approach has a number of known limitations. Publicly available datasets, for example, often contain no business-related information, e.g., about prices or profits, making the assessment of the business value of a recommender difficult. Additionally, it is, unfortunately, often not clear under which circumstances the data was collected.
For datasets that contain user interaction logs (e.g., e-commerce transactions or listening logs on music sites), the data can be biased in different ways \cite{Joachims2017UnbiasedLW,KirshenbaumForbes}, e.g., by an already existing recommender on the site, by the order in which items were presented, or by promotions that were launched during the data collection period. Evaluations that are based on such logs might lead to wrong or biased conclusions, such as an over- or underestimation of a recommender's effectiveness.

In recent years, different efforts have been made to deal with these problems. One line of research proposes alternative evaluation measures, which take into account biases, e.g., by considering that observations or ratings are not missing at random \cite{Steck:2010:TTR:1835804.1835895}. An alternative approach was more recently proposed in \cite{Debiased2019}, where the goal is to obtain a more unbiased dataset through sampling.

Probably the most well-explored area in this context is that of unbiased offline evaluation mechanisms for contextual bandit or reinforcement learning approaches. In a comparably early work, the authors of \cite{Li:2011:UOE:1935826.1935878}, for example, propose a novel \emph{replay} protocol instead of using a more common simulation-based approach. Specifically, the protocol considers biases that can be found in logs when explore-exploit recommendation strategies are used, e.g., in domains like news recommendation, where new items constantly appear and have to be explored. In their analyses, the authors in particular found that the evaluation method is effective in terms of providing unbiased estimates of relevant business measures such as the total payoff. In recent years, various approaches have been proposed to use data that was collected with a given \emph{policy} (recommendation strategy) to assess the true effects of another policy.\footnote{Technical terms in this context are \emph{off-policy evaluation} and \emph{counterfactual estimation}.} The ultimate goal of such approaches is often to support \emph{``offline A/B testing''} \cite{Gilotte:2018:OAT:3159652.3159687}, so as to avoid costly and potentially risky field tests. Generally, the research community in this field is highly active, with recent works that, for example, address the issue of user preferences changing over time \cite{Jagerman:2019:PCM:3289600.3290958} or propose new metrics that make better use of available log data \cite{Recap2019}.

\subsubsection{Limitations of Accuracy as a Proxy for Business Value}

Besides problems related to the underlying data, it is often not fully clear to what extent the abstract accuracy measures used in typical offline experiments (like RMSE\footnote{
Generally, the prediction of ratings is a traditional computational task in the recommender systems literature, but is nowadays considered less relevant in practice, because explicit ratings are rare in many applications. Therefore, research based on implicit feedback signals, i.e., the users' observed behavior, has become predominant and almost all of the works examined in our survey are rather based on implicit feedback than on expressed ratings. This can, in general, be seen as a positive development, as implicit feedback can often give clearer insight into a recommender's effect on the customer's consumption behavior than abstract rating values.}, precision, or recall) are correlated with the business success of a recommender. Intuitively, having an algorithm that is able to better predict than another whether a user will like a certain item should lead to better or more relevant recommendations. However, if this leads to increased business value, is not always clear. Users might, for example, rate items highly when they try to give an objective opinion online. Yet, they might not want to purchase similar items in the future, because, subjectively, the item does not satisfy them. As a result, it might have been better to make a riskier recommendation, which might lead to additional~sales.

Gomez-Uribe and Hunt~\cite{Gomez-Uribe:2015:NRS:2869770.2843948} discuss the general challenges of offline experiments at Netflix and, as mentioned above, conclude that they are not always indicative of online success. In fact, a number of research works exist that compare algorithms both in field tests and in offline tests or user studies. Surprisingly, in the majority of these attempts, the most accurate offline models did neither lead to the best online success nor to a better accuracy perception \cite{Rossetti:2016:COO:2959100.2959176,Cremonesi:2012:IPP:2209310.2209314,
Garcin:2014:OOE:2645710.2645745,Maksai:2015:POP:2792838.2800184,DBLP:conf/ercimdl/BeelL15,JannachLerche2017,Ekstrand:2014:UPD:2645710.2645737,McNee:2002:RCR:587078.587096}.
Only a few works report that offline experiments were predictive of what was observed in an A/B test or a user study, e.g., \cite{Brovman:2016:OSI:2959100.2959166,10.1007/978-3-642-40477-1_21,Kamehkhosh2017}. This particular problem of offline experiments is, however, not limited to recommender systems and can also be observed in other application areas of machine learning, e.g., click prediction in advertising. The authors of \cite{Yi:2013:PMP:2487575.2488215}, for example, discuss problems of measures such as the AUC and propose an alternative evaluation approach.
Overall, it remains to be shown through more studies that small improvements in offline accuracy measurements---as  commonly reported in academic papers---actually have a strong effect on business value in practice. This is in particular important as studies show that even algorithms with similar offline accuracy performance can lead to largely different recommendations in terms of the top-n recommended items \cite{JannachLercheEtAl2015}. The work in \cite{Ekstrand:2014:UPD:2645710.2645737} also indicates that methods that lead to good RMSE values can result in recommendations that are perceived to be rather obscure by users (even though they might actually be relevant). This might, in fact, be a reason why Netflix uses ``a pretty healthy dose of (unpersonalized) popularity'' in their ranking method \cite{Gomez-Uribe:2015:NRS:2869770.2843948}.
Note that we generally do not rule out that established metrics for offline evaluation like NDCG or MRR \emph{can} be predictive of the business value of a deployed algorithm. This correspondence, however, has to be validated for each particular business case and cannot be assumed in general.\footnote{In \cite{ludewigjannach2019radio}, some correspondence of precision and recall with user-perceived quality levels could be established. However, one of the algorithms that was compared in this user study performed very poorly in an offline measurement, but received acceptance scores as high as other algorithms by users in a survey.}

On a more general level, most works on recommender system \emph{algorithms} can be considered as research in applied machine learning. Therefore, they can suffer from certain limitations of today's research practice in this field \cite{2018arXiv180703341L,dacremaetal2019} and in particular from the strong focus of aiming to ``win'' over existing methods in terms of individual (accuracy) measures \cite{Lin:2019:NHC:3308774.3308781,DBLP:journals/corr/abs-1206-4656}. In this context, it can happen that improvements that are reported in the academic literature over several years ``don't add up'', as shown already in 2009 in \cite{Armstrong:2009:IDA:1645953.1646031} for the Information Retrieval domain. Similar observations were made more recently for improvements that were attributed to deep learning techniques, where indications were found that sometimes long-established and comparably simple methods, when properly tuned, can outperform the latest algorithms based on deep learning techniques \cite{Lin:2019:NHC:3308774.3308781,LudewigJannach2018,dacremaetal2019}.

\subsubsection{Beyond-Accuracy Measures: Novelty, Diversity, Serendipity, and Coverage}
In the area of recommender systems, it has been well established for many years that optimizing for prediction accuracy ``is not enough'' \cite{McNee:2006:AEA:1125451.1125659} and that several other quality factors should be considered in parallel. Recommendations should, for example, have some level of novelty to help users discover something new or should be diversified to avoid monotonous recommendations of items that are too similar to each other. Correspondingly, a number of metrics were proposed to measure these quality factors, e.g., by quantifying diversity based on pair-wise item similarities or by determining novelty based on item popularity \cite{Vargas:2011:RRN:2043932.2043955,Kaminskas:2016:DSN:3028254.2926720}. Likewise, various algorithmic proposals were made to balance accuracy with these quality factors on the global \cite{Zhang:2008:AMI:1454008.1454030,Zhou4511} or individual \cite{JugovacJannachLerche2017eswa} level, as there usually exists a trade-off situation.
From a technical viewpoint, in particular contextual-bandit approaches, as mentioned above, are often used to deal with the \emph{explore-exploit} problem, i.e., to find a balance between recommending assumedly relevant content and exploring the relevance of novel items. Specifically, such bandit-based recommenders can be designed to randomly serve a small amount of novel content to users, e.g., newly published articles in the news domain \cite{Li:2011:UOE:1935826.1935878}, and use the feedback received on these novel recommendations to learn if this content should also be recommended to others.
Such approaches not only increase the chances of novel items being presented to users, they also represent a possible countermeasure against feedback loops, which can occur when recommender algorithms continue to promote the same limited set of items resulting in a ``rich-get-richer'' or blockbuster effect. Note that besides novelty, bandit-based approaches were also designed to boost other quality factors like user coverage or diversity \cite{Rahman2018,doi:10.1137/1.9781611973440.53,Radlinski:2008:LDR:1390156.1390255}.

In the real-world applications described in Section \ref{sec:business-value}, different measurements are directly or indirectly related to metrics that measure quality factors beyond accuracy. Catalog coverage is, for example, considered as a direct quality measure in the video streaming domain. Furthermore, being able to make recommendations that are both novel, diverse, and relevant can help to better leverage the long tail item spectrum, to point consumers to other parts of the catalog and thereby increase profit or sales diversity in e-commerce settings. Similarly, serendipitous and diversified recommendations might often lead to higher levels of user engagement and customer retention in other domains.

In some ways, beyond-accuracy metrics therefore have the potential to narrow the gap between offline experimentation and field tests, as they enable a finer-grained and multi-faceted assessment of the recommendations that are generated by an algorithm \cite{said2012}. More research is, however, still required. For example, for many beyond-accuracy measures used in the literature, e.g., for intra-list diversity \cite{Ziegler:2005:IRL:1060745.1060754}, it is not always fully clear to what extent they correlate with the actual user perception. Similar challenges exist for novelty and serendipity measures. On the other hand, little is known about how diversity and novelty aspects are considered within algorithms in real-world applications. In the studies reviewed in this survey, we can observe that business-oriented measurements are made that have a strong relation with beyond-accuracy quality factors, but usually no details are provided on how, e.g., diversification is actually ensured algorithmically.

\subsubsection{Predicting Effects and Business Value}
Ultimately, the holy grail in the context of offline experimentation is to find proxy measures that correlate well with the different forms of business success measures. So far, it seems that achieving this goal remains challenging for different reasons. On the one hand, practical success measures are often very specifically tailored to the application domain or even to the business model. On the other hand, academic researchers usually aim to abstract from domain specifics and to develop generalizable solutions that are applicable to many~domains.

Currently, our knowledge is limited to certain general tendencies of algorithm families. Content-based techniques, for example, can, by design, lead to limited discovery effects, as they aim to retrieve the most similar items from the catalog. Collaborative filtering techniques, on the other hand, are often more suited to make serendipitous recommendations, but these recommendations might also be more ``risky''. Furthermore, within the family of collaborative approaches, there are some techniques like Bayesian Personalized Ranking \cite{Rendle2009} that have a tendency to recommend already popular items, whereas certain matrix factorization techniques also recommend niche or almost obscure items \cite{Ekstrand:2014:UPD:2645710.2645737}. More research in terms of understanding ``what recommenders recommend'' \cite{JannachLercheEtAl2015} and how a recommender might affect consumer behavior and business value is therefore needed. Characterizing an algorithm only with abstract quality measures---even if including beyond-accuracy measures---seems insufficient as long as the implications for practical applications are not considered. Generally, this calls for a richer methodological repertoire, which should, for example, also consider simulation experiments and alternative ways of assessing business value, see also \cite{JannachAdomaviciusVAMS2017}.

\section{Implications}
\subsection{Implications for Businesses}
Our survey of real-world deployments of recommender systems in Section \ref{sec:business-value} shows that there are many cases where such systems substantially contribute to the success of a business. These systems either help to increase revenue or profit directly, or they lead to indirect positive effects such as higher user engagement, loyalty, and customer retention. Overall, there is ample evidence that recommenders can have a strong impact on user behavior and can therefore represent a valuable tool for businesses, e.g., to steer consumer demands. Nonetheless, the expected size of the impacts depends strongly on the specific situation and the used measurements. While there are reports that recommenders lead to 35\,\% of additional revenue through cross-sales in the case of Amazon, direct revenue increases are more often reported to lie between one and five percent, which can also be substantial in absolute numbers.

Generally, our review shows that measuring the value of a recommender system is not trivial. Even when revenue or profit can be captured directly in A/B tests, there might be longitudinal effects that are difficult to assess in advance. In many cases, however, indirect measurements have to be used, e.g., by approximating customer retention through user engagement. In such situations, it is important to ensure that the underlying assumptions are thoroughly validated in order to be certain that we do not optimize for the wrong objective. Overall, the \emph{choice} of the evaluation criterion is one of the most crucial aspects in practical deployments of recommenders. Click-through rates are often used as the measure of choice---partly because it is easy to acquire---but many reports show that CTR measurements can be misleading and do not actually capture the business value well. To avoid such problems, it is therefore necessary to make sure that the strategic or operational objectives of the business are considered when designing a recommendation algorithm and when evaluating its effect, e.g., by using the purpose-oriented framework from \cite{JannachAdomavicius2016}.

\subsection{Implications for Academic and Industrial Research}
Our work also has implications for academic and industrial research. The surveyed literature indicates that substantial improvements in terms of business value can be obtained when alternative or improved algorithms are put into production. Often, however, these improvements are achieved by applying an alternative strategy (e.g., personalized vs.~non-personalized or content-based vs.~collaborative vs.~hybrid). Only in fewer cases, smaller variations of existing approaches are reported to lead to relevant impacts. Such smaller variations, e.g., in terms of the change of the loss function of a machine learning approach, are, however, very common in academic research, and it remains particularly unclear if marginal improvements on abstract measures like RMSE translate into more effective recommendations. As stated above, it has been argued that prediction accuracy is only one of several factors that determine a recommender system's effectiveness \cite{McNee:2006:AEA:1125451.1125659}. User interface design choices, in contrast, can have a much larger impact on the success of a recommender than even major algorithmic changes \cite{Garcin:2014:OOE:2645710.2645745} and should therefore be more in the focus of academic research~\cite{Konstan2012,Jannach:CommACM2016:UserControl}.

Likewise, it is undisputed in the academic literature that system-provided \emph{explanations} can have a major impact on the effectiveness of a recommender systems, see \cite{NunesJannachUmuai2017,Tintarev2012,Friedrich_Zanker_2011}. In this context, explanations can have different desirable effects. In the short term, they can help users make better decisions, make decisions faster, or persuade users to choose a certain option. In the long term, explanations are considered as an important means for trust-building. Interestingly, explanations were not prominently featured in most of the research articles that we surveyed. Therefore, we see many open opportunities in particular for industrial researchers to leverage the immense potential of explanations, e.g., in the form of persuasive cues \cite{PersuasiveRS2013} or for establishing long-term positive customer relationships.

Another avenue for future research lies in the consideration of the impact of recommender systems for different stakeholders. Current research focuses mostly on the consumer perspective, but in reality there can be a trade-off between the objectives of consumers, the recommendation platform, manufacturers, retailers, and service providers \cite{Abdollahpouri:2017:RSM:3079628.3079657,JannachAdomaviciusVAMS2017,JannachAdomavicius2016}. Academic papers, for example, rarely focus on questions such as how retailers can use recommendations to persuade users to buy more expensive items without losing their trust or how item manufacturers can be harmed by biased recommendations strategies.

Despite their limitations, offline evaluation procedures and abstract, domain-independent computational measures will remain relevant in the future to compare different algorithms. However, a number of research opportunities also exist in this context, e.g., in terms of the development of new offline evaluation procedures that lead to a more realistic assessment of the value of different recommendation strategies \cite{JannachImpactRSTowards2019}. Following the argumentation from \cite{2018arXiv180703341L}, researchers should focus more on investigating \emph{why} a given strategy led to certain effects than on merely reporting \emph{how} they obtained an improvement.
Consider, for example, that a new algorithm leads to higher \emph{recall} values in an offline experiment, which could be a desirable property for the given system. However, these higher recall values could be the result of an increased tendency of the new algorithm to recommend mostly popular items \cite{Jannach2015}. Such popularity-biased recommendations can also be undesirable from a business perspective because they limit discovery. In contrast, recommending too novel or unpopular items might be similarly detrimental to the user's quality perception of the recommender~\cite{Ekstrand:2014:UPD:2645710.2645737}. Overall, it is important to consider such underlying, generalizable theories from the literature as well as domain specifics when analyzing the outcomes of offline experiments.

Another, somehow surprising observation of our review was that we could not identify any research work that aims to assess the quality perception and helpfulness of a deployed recommender system through user satisfaction and user experience surveys. Such surveys are a very common instrument in practice to obtain feedback by real users and to improve the quality of a given service or website. Differently from the field of Computer Science, which dominates parts of the research landscape, surveys are a relatively common tool in Information Systems research to identify factors that contribute to the users' satisfaction, see, e.g., \cite{KARAHOCA2010224, ilias1970end,Jang:2016:UUF:2932206.2932207,doi:10.1089/cpb.2007.0117}. In many cases, such surveys are based on standardized questionnaires---based on factors such as information accuracy, ease of use, or timeliness of the results---that aim to identify the strengths and weaknesses of a proposed system that might affect its user experience \cite{doll1988measurement,736098220020901,doi:10.1111/j.1540-5414.2005.00076.x}. Clearly, while such surveys do not allow us to directly measure business value, they can be valuable indicators for the acceptance of a system and for possible ways of improving the service. The lack of industrial reports on the outcomes of such surveys might be caused by several reasons, e.g., that companies do not want to reveal challenges they faced when iteratively improving the system. Sharing the main insights from such surveys might furthermore mainly help competitors in the market.
We, however, believe that such surveys represent a promising tool for researchers to understand the usefulness of recommender systems in practice.

\section{Conclusion}
Our literature survey shows that recommender systems are one of the main success stories of artificial intelligence and machine learning in practice, often leading to huge benefits for businesses. Despite their success, there are still many opportunities for future research, which however often seems hampered by today's predominant research approaches in academia.

The ultimate solution to many open issues might be to conduct more large-scale field tests in the context of industry-academia partnerships in the future. While it is difficult to achieve this long-term goal immediately, there are a number of opportunities identified throughout the paper that could help us to advance our field incrementally. As an alternative to individual cooperations with industry, public competitions could also serve as field tests, such as the CLEF NewsREEL\footnote{\url{http://www.clef-newsreel.org/}} challenge, where recommendations generated by the participating academic teams are displayed to real users.

Besides field tests, we also see a strong potential to advance the field by putting more emphasis on user-centric and impact-oriented experiments and a richer methodological repertoire than we see today. Furthermore, there are still numerous opportunities to improve our current offline experimentation approaches. These include the increased adoption of multi-dimensional evaluation approaches, the consideration of generalizable theories when assessing experimental outcomes, and the use of alternative evaluation methods, e.g., based on simulation approaches. Given the links between academia and industry that are already established today, we can also expect that more real-world datasets are published for research in the future, in particular ones that contain business-related information.

\bibliographystyle{abbrvnat}
\bibliography{literature}
\end{document}